# Growth of high quality large area MgB$_2$ thin films by reactive evaporation

Brian H. Moeckly and Ward S. Ruby

Superconductor Technologies, Inc., 460 Ward Drive, Santa Barbara, CA 93111

Email: bmoeckly@suptech.com

Abstract

We report a new *in-situ* reactive deposition thin film growth technique for the production of MgB$_2$ thin films which offers several advantages over all existing methods and is the first deposition method to enable the production of high-quality MgB$_2$ films for real-world applications. We have used this growth method, which incorporates a rotating pocket heater, to deposit MgB$_2$ films on a variety of substrates, including single-crystalline, polycrystalline, metallic, and semiconductor materials up to 4" in diameter. This technique allows growth of double-sided, large-area films in the intermediate temperature range of 400 to 600 °C. These films are clean, well-connected, and consistently display $T_c$ values of 38 to 39 K with low resistivity and residual resistivity values. They are also robust and uncommonly stable upon exposure to atmosphere and water.

Since the discovery of superconductivity in $MgB_2$,[1] there has been marked progress in the ability to grow thin films of this material by the techniques of PLD,[2] [3] MBE[4-7] and HPCVD.[8] However, the efficacy of most of these methods is deleteriously affected by the particular challenges of $MgB_2$ film growth, meaning they must employ sub-optimal growth temperatures or undesirable *ex-situ* or *in-situ* annealing procedures. Numerous difficulties thus remain regarding the routine deposition of completely *in situ*, smooth, robust, high-quality films which are of interest for applications. In addition, the deposition method should also be compatible with multilayer deposition, double-sided deposition, and the growth of films on large-area substrates of technological interest. We have developed a growth technique that promises to meet these requirements.

The primary difficulties with $MgB_2$ thin film deposition center on Mg. This element has a vapor pressure that is many orders of magnitude higher than B, and thus a relatively high pressure of Mg is needed to ensure phase stability of $MgB_2$ at the deposition temperature. Mg also reacts strongly with oxygen which can therefore contaminate the films and adversely affect their superconducting properties. Lastly, the sticking coefficient of Mg is relatively low.[9] These properties pose difficulties for most physical vapor deposition growth techniques. For example, PLD and MBE are limited to low growth temperatures because of the difficulty of providing sufficient Mg flux to grow epitaxial $MgB_2$ films at high temperatures. Thus it is also difficult to control the growth rate and the relative Mg and B fluxes to routinely achieve the correct phase. Moreover the low growth temperature of these films often leads to sub-optimal properties: the superconductive transition temperature $T_c$ is usually lower than the bulk value and the resistivity is usually higher. In addition, the films are often not dense or fully connected.[10]

These problems are tempered somewhat by properties that are favorable for film growth. One is that the growth of $MgB_2$ appears to be kinetically limited,[11] wherein the vapor pressure of Mg over $MgB_2$ is far lower than over Mg metal, so that Mg does not escape the $MgB_2$ film once it has formed at a reasonable growth $T$. The growth of $MgB_2$ also appears to be adsorption limited, so that as long as sufficient Mg vapor is present the $MgB_2$ phase will form automatically and additional Mg will not be incorporated into the film.[12] The growth technique of HPCVD makes use of these properties by providing Mg vapor near the substrate and introducing B by a carrier gas. This technique has produced very clean, epitaxial films with the lowest reported residual resistivity values and highest $T_c$ values.[8] However, the technique is currently incompatible with *in situ* multilayer deposition and is limited to high-temperature growth of small, single-sided samples on a limited number of single-crystalline substrates.

We have developed a completely *in-situ* technique requiring no post annealing that directly addresses the difficulties of $MgB_2$ film growth while capitalizing on the advantages. The technique utilizes a rotating pocket heater of a type similar to that developed for deposition of large-area HTS[13, 14] and other oxide[15, 16] thin films. Such a heater is routinely employed for high-throughput production of (RE)BCO films. The heater contains a rotating platter that holds the substrates by contacting them only at the outside edge and spins them at several hundred rpm through a quasi-black-body radiative oven. During about a third of one rotation cycle, the substrates are exposed to the vacuum chamber via a window and hence to the evaporated flux of metals and rare earths. In the usual method, the substrates then pass through the heater and are exposed to a pocket of oxygen gas where oxidation occurs. We have made a crucial modification to this technique by introducing one of the metallic constituents as a vapor directly into the pocket of the heater rather than co-evaporating it from a standard cell, boat, or crucible. For the growth of MgB$_2$ films, we thus introduce Mg vapor only into the interior of the heater as shown in the inset of Fig. 1. The Mg vapor is relatively well sealed inside the heater pocket by means of a small gap between the platter and the heater body, and our single B e-beam source is therefore free to operate in a vacuum environment. This approach offers several advantages. The required relatively high pressure of Mg is provided locally near the substrates, while deposition of B or other materials can take place in vacuum using usual PVD techniques. We have also found that UHV conditions (used to avoid incorporation of impurities, namely oxygen in this case) are not necessary in order to obtain high quality MgB$_2$ films. We speculate that excess Mg getters any residual oxygen in the pocket; this MgO is then not incorporated into the films at the high temperatures but rather escapes from the heater and adheres to cold surfaces elsewhere in the vacuum system, thereby avoiding contamination of the films. The important point is that any impurities in the system are not present during reaction, which takes place only inside the heater pocket.

Since the Mg vapor is introduced into the heater via an external heated cell, the temperature of the Mg is largely independent of the substrate temperature, unlike embodiments of the HPCVD method. In addition, our technique enables double-sided deposition simply by flipping the wafers over between runs and growing on the other side; the heater design allows contactless heating and Mg does not escape from the first MgB$_2$ film during the growth of the second due to the kinetic barrier. Our current 5"-diameter heater can accommodate a single 4" wafer or a multitude of small substrates in combination with up to three 2" wafers. Thus we can combine several substrates in a single run, which allows great efficiency in process development. And because the heater approximates a blackbody radiator, different substrate materials can be incorporated simultaneously even if they have different absortivities. A larger heater allowing deposition on larger or a greater number of substrates should be readily achievable when needed for future applications.

The heater temperature we have employed so far has been between 400 and 600 °C. This is an intermediate range that is higher than that used for MBE (limited to ~300 °C), and is below that used for optimal films deposited by HPVCD (~700 °C). Our typically substrate rotation speed is 300 rpm, and we evaporate B at about 1 Å/sec, which determines the growth rate of our films. We have not yet attempted higher rates. The background pressure in our vacuum chamber is in the range of $10^{-7}$ to $10^{-6}$ Torr, with an oxygen partial pressure during growth of typically ~$10^{-8}$ Torr. We have grown films from 150 to 700 nm thick.

Fig. 1 displays a typical $r(T)$ characteristic of our MgB$_2$ films grown on single crystal substrates. This 0.5-micron-thick film was grown at 550 °C on *r*-plane sapphire, which is not lattice-matched to MgB$_2$. Nevertheless, the resistivity at 300 °C is about 10 µΩ cm, which approaches the best bulk and single crystal values.[10] Moreover, the difference in resistivity ?$r$ between 300 K and 40 K is about 8 µΩ cm. Indeed, all of our films display ?$r$ values ranging from 7 and 9 µΩ cm, which is also in agreement with the bulk value and indicative of dense, well-connected films.[10] The main difference between our various films is the value of the residual resistivity, which can be estimated by the value of the resistivity at 40 K, equal to 2.6 µΩ cm in Fig. 1. This is a low value for MgB$_2$ thin films, bettered only by HPCVD values, which can be a fraction of a µΩ cm. Since the bulk of the grains in our films appear to be clean as evinced by the low values of ?$r$, it is possible that this residual resistivity is due to scattering at the grain boundaries. This difference in $r$(40 K) between our films and the HPCVD films may hence be because our grain size is about an order of magnitude smaller than in those films (~100 nm vs. ~1 µm).

Our $T_c$ values are routinely 38 to >39 K, on par with the bulk value. The transition to the superconducting state is sharp, with widths $?T_c$ of ~0.2 K.[17] These observations indicate that the films are clean. We have deposited similarly high quality films on a number of single crystal substrates, including *r*-plane sapphire, *c*-plane sapphire, *m*-plane sapphire, 4H-SiC, MgO, $LaAlO_3$, $NdGaO_3$, $LaGaO_3$, LSAT, $SrTiO_3$, and YSZ. It is clear that a good lattice match is not required in order to obtain excellent $MgB_2$ films. TEM studies including selected area electron diffraction indicate that our films are epitaxial and appear to be well-oriented regardless of the substrate employed,[18] having a columnar grain structure and in-plane registration of the grains. However, in most cases it is difficult to easily observe the *c*-axis lattice peaks in cursory x-ray diffraction scans, perhaps due to the small grain size of our films. In any case, the diffraction peaks of $MgB_2$ are known to be weak.

This growth technique is not limited to single-crystal substrates. For example, Fig. 2 displays the *r*(*T*) curve of an $MgB_2$ film deposited on a polycrystalline alumina substrate. In this case as well, low*r*, low ?*r*, and low residual resistivity values are obtained along with the bulk $T_c$ value. The technique appears to enable deposition of high-quality films on any substrate for which there is not a chemical reaction with Mg or B. As a further example, we have also introduced standard stainless steel coil shim stock into our process. We cleaned this flexible steel in solvents, but did not attempt to polish it. Nevertheless, we readily obtained films on this crude substrate that displayed low values of resistance and moderately high $T_c$ values, as shown in Fig. 3. This result readily demonstrates the utility of this growth technique for applications such as flexible interconnects and coated conductors.

We have also deposited $MgB_2$ on Si. Since $MgB_2$ reacts strongly with bare Si, even at 400 °C, a suitable buffer layer is necessary when using Si substrates. We obtained several Si substrates buffered with various oxides and nitrides and determined their suitability for $MgB_2$ deposition. These buffer layers were between 100 and 400 nm thick and included $SiO_2$, $SrTiO_3$, MgO, $TiO_2$, $Al_2O_3$, AlN, $Ta_2O_2$, $Nb_2O_5$, $ZrO_2$, $SiN_x$, and $Si_3N_4$. With the exception of $SiO_2$, which reacted strongly, in most cases we obtained good results. Fig. 4, for example, displays the $r(T)$ curve for $MgB_2$ deposited on Si with a buffer layer of $Si_3N_4$.

A typical AFM scan of the surface of our films is shown in the inset of Fig. 4. We usually observe small, conical-shaped grains with a diameter of 100 to 200 nm and RMS surface roughness values of 1 to 5 nm for films that are ~500 nm thick. The films are thus suitable for device and multilayer applications. The surface morphology depends somewhat on which substrate is used, and smoother films are typically obtained for thinner films grown at lower temperature. XPS studies of the surface of our films indicate the existence of a thin layer (2-3 nm) of Mg and B oxides.[19] This oxide layer appears to be very stable: the XPS measurements were made several months after the films were deposited, and the surface layer remained thin and was not degraded. N-I-S point-contact spectroscopic measurements of our films grown on $r$-plane sapphire substrates show tunneling $I$-$V$ characteristics with clear evidence of both superconducting gaps, indicating that the oxide layer acts as a good tunnel barrier.[20] Moreover, these measurements are also reproducible over many months, again indicating the excellent surface stability of our films.

The films themselves also appear to be relatively stable, in contrast to many reports for $MgB_2$. We have processed them using standard photolithography and inert ion etching, with exposure to solvents and deionized water during various stages of processing. The resistivity and $T_c$ value measured across a 10-micron-wide line so patterned yielded similar values to the bulk film, indicating that the films are dense on this size scale and that they are robust to routine processing procedures. Furthermore, we have submerged our films directly into DI water for various periods. The main effect appears to be a slow etching of the film at a rate of ~35 Å/hour. After 24 hours in water, the 38.9-K $T_c$ value of the test film remained unchanged, and the room-temperature resistivity value increased slightly from 12 to 14 $\mu\Omega$ cm. After 42 hours, $T_c$ was still high at 38 K, though the resistivity increased further to 29 $\mu\Omega$ cm. After ~60 hours, this film had indeed disintegrated, but this brief experiment indicates that our films are likely to be sufficiently stable for normal applications. Indeed, the $r(T)$ characteristics remained unchanged for films that have been stored for many months in an $N_2$ environment. Moreover, the RF properties of one of our patterned films were unchanged upon remeasurement after storage for a year in a desiccator. [21]

We measured the critical current across the above-mentioned 10-micron-wide line patterned in one of our films on an alumina substrate. The $J_c$ value reached $10^6$ A/cm$^2$ at 35.5 K. The cleanliness of our films has been verified by RBS measurements that indicate a lack of oxygen incorporation to less than the detectable level of ~1 %.[22] Values of $R_s$ of our films are the lowest reported in the literature.[23] Lastly, the $H_{c2}$ values of our films are also very low, again indicating that they are clean.[24] These low $H_{c2}$ values can be significantly raised by C doping [24] or by ion damage.[25]

We have deposited films on substrates ranging in size from 0.25" square to 4" in diameter. We have grown on 2"-diameter substrates of LAO, MgO, and sapphire, and we have deposited on 4" substrates of sapphire and Si buffered with $Si_3N_4$. The inset of Fig. 2 shows an $MgB_2$ film deposited onto a 4" *r*-plane sapphire substrate. These films are uniform over the area by visual inspection and are also of uniform thickness. We subsequently diced these wafers into $cm^2$ pieces and measured their $T_c$ values, which were uniformly high.

In conclusion, we have demonstrated a new method for completely *in situ* deposition of $MgB_2$ films. This reactive evaporation technique allows growth of high quality, large area $MgB_2$ films on a variety of substrates of technological interest and is compatible with standard physical vapor deposition protocols. Thus the method should be easily extendable to completely *in situ* deposition of multilayer structures. The films produced by this method are smooth, stable, and have high $T_c$ and $J_c$ values and low resistivity values. This method can thus enable the development of films for many potential applications. Moreover, we anticipate extending this versatile and effective reactive deposition technique to the growth of other compounds of technological importance that are also very challenging to produce in thin film form by alternative growth techniques.

ACKNOWLEDGEMENTS

We are grateful for the technical assistance of K. Von Dessonneck and for the buffered Si substrates provided by R. Buhrman, X.-F. Meng, and T. Van Duzer. We thank J. Rowell and R. Buhrman for advice and encouragement. This research was supported in part by the Office of Naval Research, Contract No. N00014-03-M-0005.

**FIGURE CAPTIONS**

FIG. 1 Resistivity versus temperature for a 550-nm-thick $MgB_2$ film deposited on $r$-plane sapphire at 550 °C. The zero-resistance $T_c$ of this film is 39.1 K. The inset shows a drawing of our heater and deposition technique.

FIG. 2 Resistivity versus temperature for an $MgB_2$ film deposited on polycrystalline alumina. This film was grown at 550 °C to a thickness of 550 nm and has a zero-resistance $T_c$ value of 39.1 K. The inset shows a photograph of an $MgB_2$ film grown on a 4"-diameter $r$-plane sapphire substrate.

FIG. 3 Resistance versus temperature for a half-micron-thick $MgB_2$ film grown on unpolished stainless steel shim stock at 550 °C. The film has a $T_c$ value of 36.5 K. A photograph of the film is shown in the inset.

FIG. 4 Resistivity versus temperature for an $MgB_2$ film grown on Si with a 100-nm buffer layer of $Si_3N_4$. The $T_c$ of this 500-nm-thick film is 39.2 K. The inset shows a typical AFM scan of an $MgB_2$ film grown on $r$-plane sapphire.

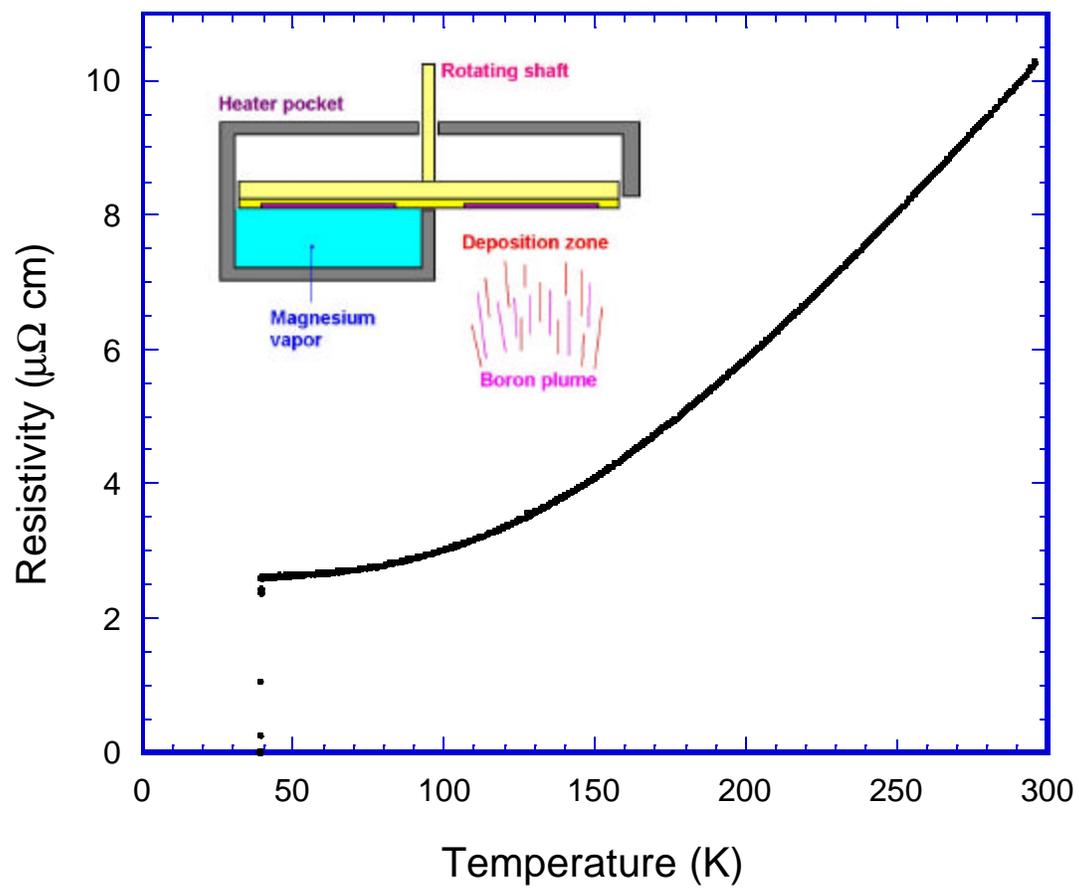

Fig. 1

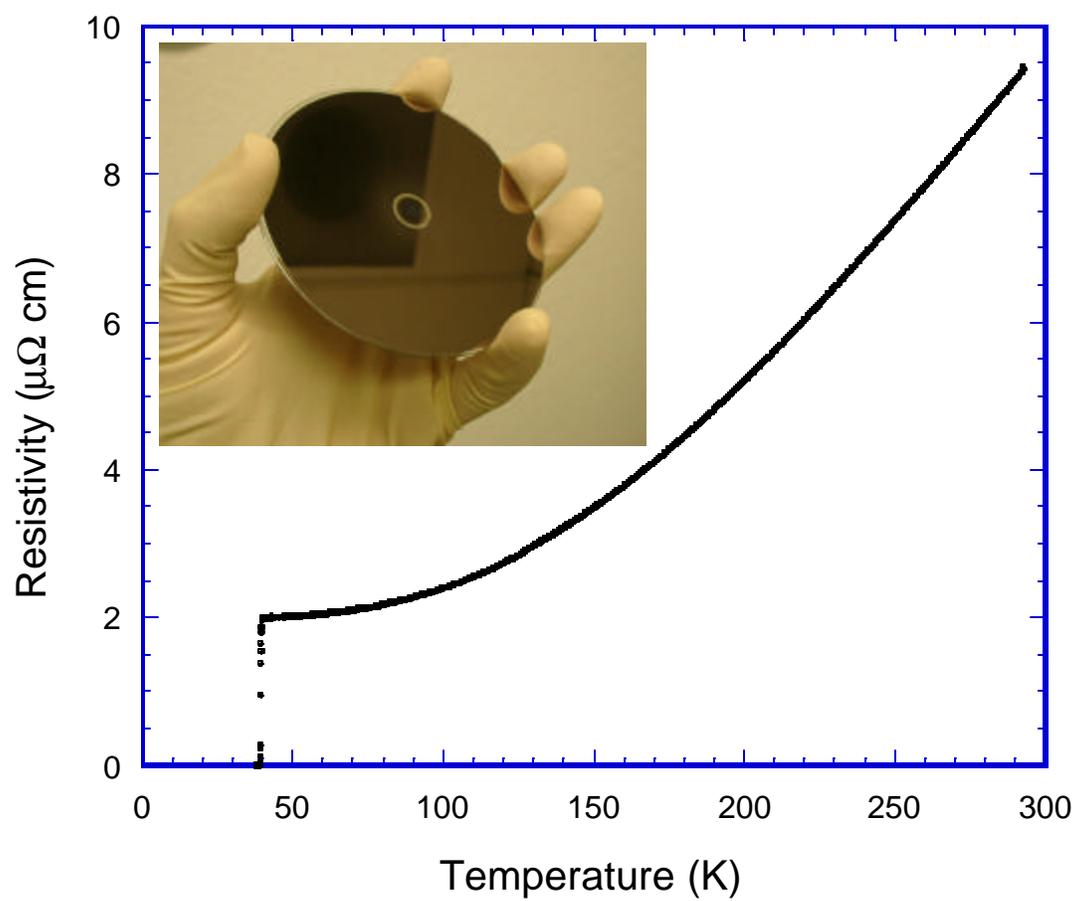

Fig. 2

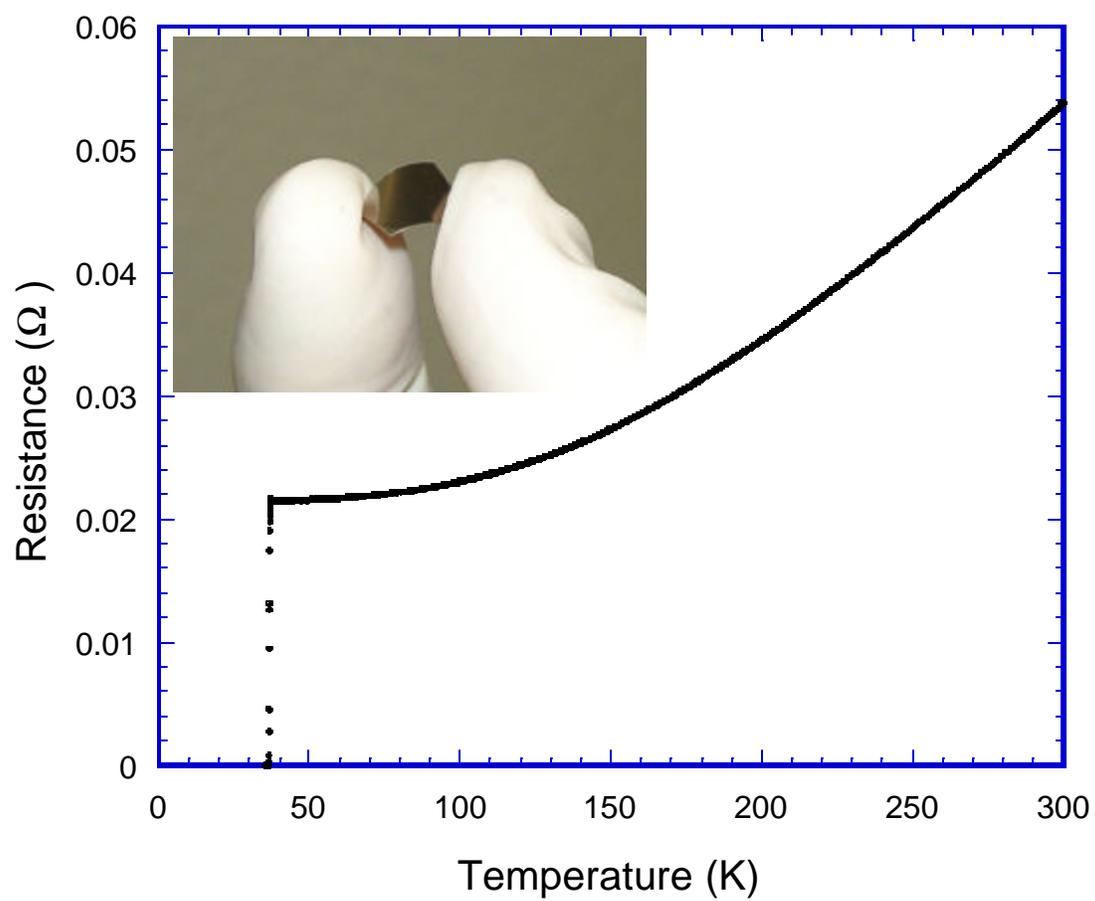

Fig. 3

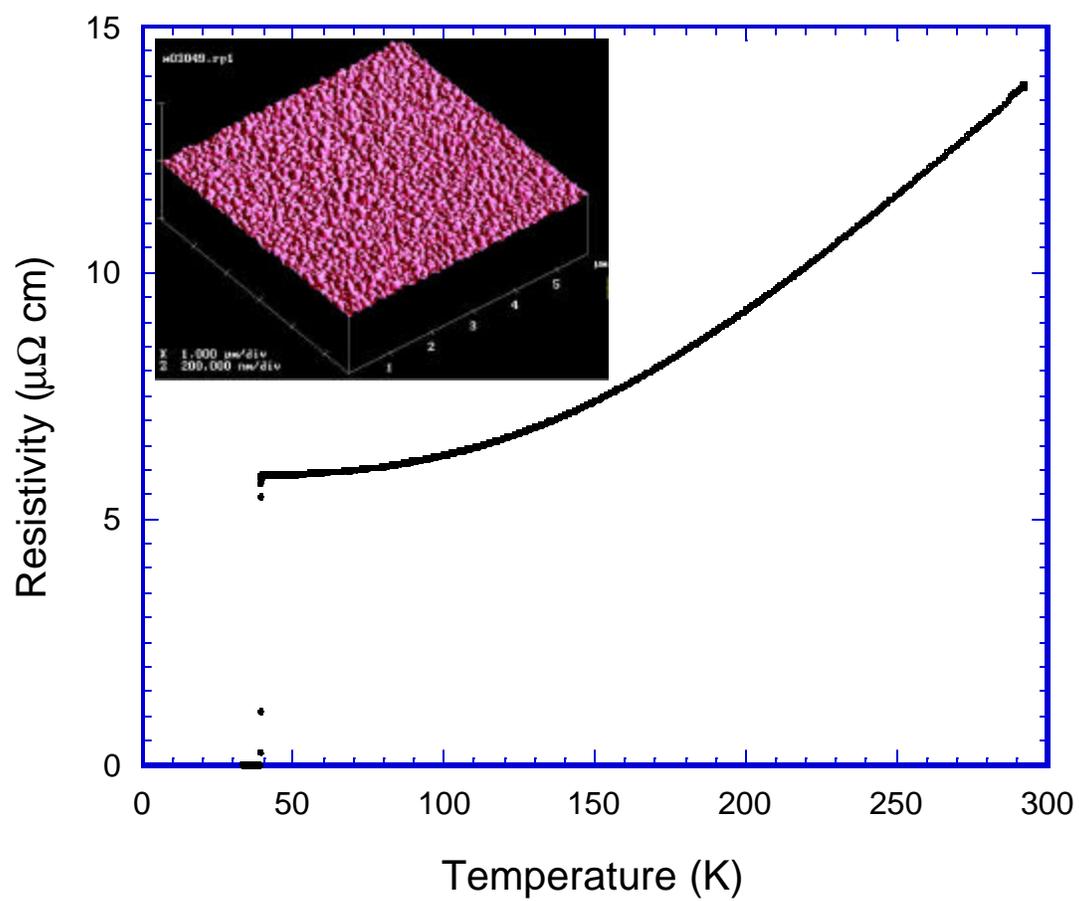

Fig. 4